\documentclass{an_2c}

\def\45{RX\,J1239.3+2431}
\def\144{RX\,J1225.7+2055}
\def\ros{{\it ROSAT }}

\begin{document}
\sloppy

\title{Testing Narrow-line Seyfert\,1 scenarios with photoionization models} 
\author{Stefanie Komossa\inst{1}   
\and  Janek Meerschweinchen\inst{2}}
\institute{
Max-Planck-Institut  f\"ur extraterrestrische Physik, D-85740 Garching, Germany; skomossa@mpe.mpg.de 
\and 
Weststrasse 19, D-3063 Obernkirchen\,2, Germany}
\headnote{Astron. Nachr. 320 (1999)}
\maketitle

\vspace*{-6cm}
\begin{verbatim}
Contribution to `Heating and Acceleration in the Universe' 
(Tokyo, March 17-19, 1999), H.Inoue et al. (eds) 
Preprint available at URL  http://www.xray.mpe.mpg.de/~skomossa
\end{verbatim}
\vspace*{4cm}

\section{Introduction}
Narrow Line Seyfert 1 (NLSy1) galaxies have recently received a lot of attention
due to their unusual optical--X-ray properties (e.g., Puchnarewicz et al. 1992,
Brandt et al. 1997, Grupe et al. 1999) which are not yet well understood.
Photoionization models of the circum-nuclear emission/absorption regions
allow to investigate
scenarios to explain the
main characteristics of NLSy1s, i.e., (i) extremely steep X-ray spectra within the \ros energy range,
(ii) narrow (FWHM $<$ 2000 km/s) Balmer lines
and (iii) weak forbidden lines except for some relatively strong high-ionization iron lines.
Here, we focus on (iii) due to the short space (for a discussion of (i) and (ii)
see Komossa \& Fink 1997a, and Komossa \& Janek 1999).
In particular, we study the influence of different EUV - soft-X-ray spectral shapes 
(a giant soft excess, a steep X-ray powerlaw, presence of a warm absorber) 
and NLR cloud properties 
on the predicted optical emission-line ratios. 
The calculations were carried out with
the code {\em{Cloudy}} (Ferland, 1993).

\section{Modeling the optical NLR emission lines }

The narrow emission lines, i.e. those originating from the narrow line region (NLR),
like [OIII]$\lambda$5007 and [OI]$\lambda$6300,
are rather weak in many NLSy1s.
Occasionally, however, fairly strong high-ionization iron lines are present.
We investigate several models to explain these
observations, starting with the assumption that the NLR is `normal'
(i.e. a Seyfert-typical one, as far as distance
from the nucleus, gas density and covering factor are concerned).

In a first step, non-solar metals abundances were employed. 
{\em Over}abundant metals (with respect to
the solar value) were shown to delay the complete removal of a BLR multiphase
equilibrium (Komossa \& Janek 1999). Due to their rather strong influence on the cooling,
metals, if overabundant, 
can lead to weaker optical line emission. 
However, we find the effect to be insufficient to explain
the observed line intensities.

As shown in Komossa \& Schulz (1997) the weak [OIII]$\lambda$5007 end of the line
correlations in the usual diagnostic diagrams for Seyfert 2s can be explained by
very steep EUV continua with $\alpha_{\rm uv-x}$ $\approx$ --2.5. Although, e.g., 
the NLSy1 \144 exhibits
a rather steep EUV spectrum (determined by a powerlaw connection between the flux at the
Lyman limit and 0.1 keV), that of \45 is very flat (Greiner et al. 1996).
 
Placing warm absorbing material along the line of sight to the NLR would make
the latter see a continuum that is only modified in the soft X-ray region, with
negligible influence on the line emission. The same holds for an intrinsically
steep X-ray powerlaw, which only leads to a slight weakening of low-ionization lines.

In case a warm absorber is present, high-ionization iron lines ([FeX] and higher)
can be produced within the warm gas itself (see Komossa \& Fink 1997a,b,c for details).
However, no one-to-one match between the observed coronal lines in the NLSy1 NGC\,4051
and those predicted to arise from the X-ray warm absorber in this galaxy was
found (Komossa \& Fink 1997a), suggesting that in general, coronal line
region and warm absorber are separate components.

In order to assess the influence of a strong EUV - soft-X-ray excess 
on optical line emission, 
we have calculated a sequence of models with an underlying mean Seyfert continuum
plus a black body of varying temperature
for a range of densities and distances of the NLR gas from the central source. 
Although the contribution of a hot bump-component can considerably strengthen the
high-ionization iron lines (like [FeX]$\lambda$6374, [FeVII]$\lambda$6087 
and  [FeXIV]$\lambda$5303), reflecting the fact that their
ionization potentials are at soft X-ray energies,
these models overpredict the [OIII] emission.

We conclude that the weakness of forbidden lines in NLSy1s must be due to an overall
lower emissivity of the NLR.
If this is caused by a shielding of the NLR from photons,
care must be taken not to boost
the low-ionization lines like [OII]. 
More likely, the region is short of gas, i.e. less of the impinging photons can be
reprocessed to line emission.
`Normal' line ratios would then leave the weak forbidden lines undetected.
For those objects with strong high-ionization (iron) lines, 
models favor the dominance of low-$r$, low-density
gas. 



\begin{thebibliography}{}
\bibitem[]{}Brandt W.N., Mathur S., Elvis M.: 1997, MNRAS 271, 958
\bibitem[]{}Ferland G.: 1993, Univ. of Kentucky, Phys. Dept, Int. Report
\bibitem[]{}Greiner J., et al.: 1996, A\&A 310, 384 
\bibitem[]{}Grupe D., et al.: 1998, A\&A 330, 25; 1999, A\&A accepted   
\bibitem[]{}Komossa S., Janek M.: 1999, A\&A, submitted
\bibitem[]{}Komossa S., Fink H.: 1997a, A\&A 322, 719 
\bibitem[]{}Komossa S., Fink H.: 1997b, in {\sl Emission Lines
            in Active Galaxies: New Methods and Techniques},
             B.M. Peterson et al. (eds), ASP conf. ser. 113, 246
\bibitem[]{}Komossa S., Fink H.: 1997c, Lecture Notes in
             Physics 487, 250; (astro-ph/9612185)
\bibitem[]{}Komossa S., Schulz H.: 1997, A\&A 323, 31
\bibitem[]{}Puchnarewicz E.M., et al.: 1992, MNRAS 256, 589 
\end{thebibliography}
\end{document}